\documentclass[doublecol]{epl2}
\usepackage[utf8]{inputenc}
\usepackage[T1]{fontenc}
\usepackage[english]{babel}

\usepackage[active]{srcltx}
\usepackage{graphicx,soul}
\usepackage[dvipsnames]{xcolor}

\usepackage{amsmath}
\usepackage{amsfonts,amssymb,amsthm}
\usepackage{mathrsfs,mathdots}
\usepackage[centercolon=true]{mathtools}
\usepackage{braket,bm}
\usepackage{microtype}
\usepackage{enumerate}

\usepackage{hyperref,booktabs,bookmark}
\usepackage{etoolbox}

\definecolor{mycolor}{rgb}{0.2,0.5,0.8}

\hypersetup{
    colorlinks=true,
    citecolor=mycolor
}

\definecolor{shadecolor}{rgb}{.8,.8,.8}

\providebool{expand}
\setbool{expand}{false}

\makeatletter
\@ifpackageloaded{bm}
{
    
}
{
    
}
\makeatother

\providecommand*{\SMbinom}[2]{\paren*{\!\begin{smallmatrix}{#1}\\{#2}\end{smallmatrix}\!}}
\providecommand*{\PMbinom}[2]{\begin{pmatrix}\!{#1}\\{#2}\!\end{pmatrix}}

\renewcommand*{\epsilon}{\varepsilon}
\renewcommand*{\theta}{\vartheta}
\renewcommand*{\rho}{\varrho}
\renewcommand*{\phi}{\varphi}

\providecommand*{\numberset}{\mathbb}
\providecommand*{\Naturals}{\numberset{N}}
\providecommand*{\Integers}{\numberset{Z}}

\DeclarePairedDelimiter{\abs}{\lvert}{\rvert}

\DeclarePairedDelimiter{\paren}{\lparen}{\rparen}

\DeclareMathOperator{\dif}{d\!}

\providecommand*{\od}[3][]{
    \ifinner
        \tfrac{\dif{^{#1}}#2}{\dif{#3^{#1}}}
    \else
        \dfrac{\dif{^{#1}}#2}{\dif{#3^{#1}}}
    \fi
}

\providecommand*{\pd}[3][]{
    \ifinner
        \tfrac{\partial{^{#1}}#2}{\partial{#3^{#1}}}
    \else
        \dfrac{\partial{^{#1}}#2}{\partial{#3^{#1}}}
    \fi
}

\providecommand*{\mvec}[1]{
    \ifinner
        \begin{psmallmatrix}#1\end{psmallmatrix}
    \else
        \begin{pmatrix}#1\end{pmatrix}
    \fi
}

\providecommand*{\Hilbert}[1][H]{\mathcal{#1}}

\providecommand*{\Tped}[1]{_{{\textup{#1}}}}

\begin{document}

\title{Discrete Feynman propagator for the Weyl quantum walk in $2+1$
  dimensions}

\author{
  G.~M.~D'Ariano\inst{1,2}\thanks{E-mail: \email{dariano@unipv.it}} \and 
  N.~Mosco\inst{1}\thanks{E-mail: \email{nicola.mosco01@ateneopv.it}} \and 
  P.~Perinotti\inst{1,2} \thanks{E-mail: \email{paolo.perinotti@unipv.it}} \and
  A.~Tosini\inst{1,2}\thanks{E-mail: \email{alessandro.tosini@unipv.it}} }
\shortauthor{
  G.~M.~D'Ariano \etal }
\institute{
  \inst{1} QUIT group,  Dipartimento di Fisica, via Bassi   6, 27100 Pavia, Italy. \\
  \inst{2} INFN Sezione di Pavia, via Bassi, 6, 27100 Pavia, Italy. 
}

%  \pacs{03.67.Lx}{Quantum computation architectures and implementations}
% \pacs{02.10.Ox}{Combinatorics; graph theory}
% \pacs{05.40.Fb}{Random walks and Levy flights}
% \pacs{03.65.Db}{Functional analytical methods}

\abstract{Recently quantum walks have been considered as a possible fundamental description of the dynamics of relativistic quantum fields. Within this scenario we derive the analytical solution of the Weyl walk in $2+1$ dimensions.  We present a discrete path-integral formulation of the Feynman propagator based on the binary encoding of paths on the lattice. The derivation exploits a special feature of the Weyl walk, that occurs also in other dimensions, that is closure under multiplication of the set of the walk transition matrices. This result opens the perspective of a similar solution in the $3+1$ case.}

\maketitle

A simple description of particles propagation on a discrete
spacetime was proposed by Feynman in the so called \emph{checkerboard
  problem} \cite{feynman1965quantum} that consists in finding a simple
rule to represent the quantum dynamics of a Dirac particle in $1+1$
dimensions as a discrete path-integral. 

The definition of a discrete path-integral is closely related to the
underlying notion of ``discrete spacetime'' and on the dynamical model
used to describe the discrete time evolution of the quantum systems.
As a consequence, in the absence of an established theory of quantum
spacetime, the formulation of a discrete Feynman propagator can be
considered within different possible scenarios.

Following the original idea of Feynman, and the subsequent progress of Refs.~\cite{jacobson1984quantum,karmanov1993derivation}, in
Ref.~\cite{kauffman1996discrete} Kaufmann and Noyes analysed the checkerboard problem, providing a solution of the finite-difference version of Dirac's equation. In
Refs.~\cite{johnston2008particle,johnston2009feynman} a
path-integral formulation for the discrete space-time is presented within the causal set approach of Bombelli and Sorkin \cite{bombelli1986quantum}, with trajectories within the causal set summed over to obtain a particle propagator.  More recently, following the pioneering papers~\cite{succi1993lattice,meyer1996quantum,bialynicki1994weyl},
the \emph{quantum walks} (QWs) have been considered as a discrete model of dynamics for relativistic particles
\cite{PhysRevA.73.054302,
Yepez:2006p4406,darianopla,bisio2013dirac,BDTqcaI,d2013derivation,arrighi2013dirac,arrighi2013decoupled,farrelly2014causal,farrelly2013discrete}.

A QW is the quantum version of a (classical) \emph{random walk} that
describes a particle moving in discrete time steps and with
certain probabilities from one lattice position to the neighboring
sites. The first QW appeared in \cite{aharonov1993quantum} where the
measurement of the $z$-component of a spin-$1/2$ quantum system, also
denoted internal degree of freedom or \emph{coin} system, decides
whether the particle moves right or left. Then the measurement was
replaced by a unitary operator on the coin system
\cite{ambainis2001one} with the QW representing a discrete unitary
evolution of a particle state with internal degree of freedom given by
the coin.  In the more general case the coin at a site $x$ of the
lattice can be represented by a finite dimensional Hilbert space
$\Hilbert_x=\mathbb{C}^s$, with the total Hilbert space of the system
given by the direct sum of all sites Hilbert spaces.

QWs provide the one-step free evolution of one-particle quantum
states, however, replacing the quantum state with a quantum field on
the lattice, a QW describes the discrete evolution of non interacting
particles with a given statistics--a ``second quantization'' of the
QW. This can be ultimately regarded as a \emph{quantum cellular
  automaton} \cite{schumacher2004reversible} that is linear in the
filed.  QWs have been largely investigated and formalized in
computer-science and quantum information
\cite{schumacher2004reversible,arrighi2011unitarity,gross2012index,ambainis2001one,
  knight2004propagating, ahlbrecht2011asymptotic} with relevant
applications in designing efficient quantum algorithms
\cite{childs2003exponential, ambainis2007quantum,
  magniez2007quantum,farhi2007quantum}.

As pointed out in Ref.~\cite{ambainis2001one} a walk can be
analyzed in two different ways. On one hand one can diagonalize the QW
in the momentum space, on the other hand one can consider a discrete
path-integral approach, expressing the walk transition amplitude to a
given site as a combinatorial sum over all possible paths leading to
that site. Within the last perspective some QWs in one spatial
dimension have been analytically solved, the \emph{Hadamard walk}
\cite{ambainis2001one}, where the Hadamard unitary is the operator on
the coin system, the \emph{coined} QWs \cite{konno2002quantum}, with
an arbitrary unitary acting on the coin space, and the
\emph{disordered} QWs \cite{konno2005path}, where the coin unitary is
a varying function of time. 

In Refs.~\cite{BDTqcaI,d2013derivation} the authors have derived the
simplest QWs in $d+1$ dimensions, $d=1,2,3$ 
that satisfy elementary symmetry
requirements and that, as a consequence, give the usual Dirac and Weyl
equations in the limit of small wave-vectors with respect to the lattice step. The small wave-vector approximation coincides with the relativistic limit if the lattice step is hypothetically assumed equal to the Planck scale. 

While in Ref.~\cite{d2014path} the discrete path-integral solution is given for the Dirac walk in $1+1$ dimensions, here we present a technique which can be used to solve the discrete path-integral for walks in dimension higher than one. First, if the transition matrices of the walk form a closed algebra under multiplication, one can split the paths connecting two arbitrary sites on the lattice into equivalence
classes according to their overall transition matrix. Upon a suitable choice of labeling, one can encode paths into binary strings, and associate  specific algebraic properties of the strings with the overall transition matrix and with the couples of lattice points connected by the path. This remarkable correspondence allows us to classify strings that connect two given points in a given number of steps, with the same overall transition matrix. Finally, by a combinatorial analysis, the number of strings in each equivalence class is counted.  In this paper we apply this approach to the Weyl walk in $2+1$ dimensions providing its analytical solution. However, the same scheme can be used in principle for any QW which allows for a classification of paths in terms of algebraic properties of the encoding strings.

The Weyl QW of Ref.~\cite{d2013derivation} describes the one-step
linear evolution of a two-component quantum field on the two
dimensional square lattice $\mathbb{Z}^2$
\begin{align*}
  \psi (x,y,t) :=\begin{pmatrix}
\psi_1(x,y,t)\\\psi_2(x,y,t)\end{pmatrix},\quad (x,y)\in\mathbb{Z}^2,\,t\in\mathbb{Z},
 \end{align*}
 where $\psi_1$ and $\psi_2$ denote the two modes of the field. Here
 we restrict to the one-particle sector and the statistics is not
 relevant, but the presented solution is straightforwardly extended to
 free multi-particle state. In the single-particle Hilbert space
 $\ell^2(\mathbb Z^2)\otimes\mathbb{C}^2$ we use the factorized basis $\ket{x}\ket{s}$, with $x\in\mathbb Z^2$ and $s=1,2$.

 The walk is a unitary operator $A$ that gives the one-step update of
 the field $\psi(t+1)=U\psi(t) U^\dag=A\psi(t)$. The evolution is
 required to be local corresponding to write $\psi(x,y,t+1)$ as a linear
 combination of the field values $\psi(x\pm 1, y\pm 1,t)$ on the nearest neighbouring sites, and translationally invariant, corresponding to a
 unitary operator of the form
\begin{align}\label{eq:Weyl}
  A &= \sum_{h}T_{h}\otimes A_{h},\qquad h=\textup{R},\textup{L},\textup{U},\textup{D}.
\end{align}
In the last equation the symbol $T_h$,
$h=\textup{R},\textup{L},\textup{U},\textup{D}$, represents the
translation operators on the square lattice, respectively in the
\emph{right}, \emph{left}, \emph{up} and \emph{down} direction, while
the $A_{h}$ are the $2\times 2$ \emph{transition matrices} of the walk
acting on the coin system.  In the Weyl case the transition matrices
are
\begin{equation}\label{eq:transition-matrices}
\begin{aligned}
    A\Tped{R} & = \frac{1}{2}
    \begin{pmatrix}
          1  & 0 \\
        -\nu & 0
    \end{pmatrix}, &
    A\Tped{U} & = \frac{1}{2}
    \begin{pmatrix}
         1  & 0 \\
        \nu & 0
    \end{pmatrix}, \\
    A\Tped{L} & = \frac{1}{2}
    \begin{pmatrix}
        0 & \nu^{*} \\
        0 &  1
    \end{pmatrix}, &
    A\Tped{D} & = \frac{1}{2}
    \begin{pmatrix}
        0 & -\nu^* \\
        0 &   1
    \end{pmatrix},
\end{aligned}
\end{equation}
with $\abs{\nu} = 1$.  

In Ref.~\cite{d2013derivation} the dynamics of the Weyl walk has been
studied in the wave-vector space. Diagonalizing the walk, and
interpreting the wave-vector $k$ as the momentum, it has been shown
that the usual Weyl equation kinematics is recovered for small momenta
($k \to 0$). This means that there exists a class of states whose walk
evolution is indistinguishable with respect to the usual Weyl equation
solutions.

Taking the initial condition $\psi(0)$, the field at time $t$ is given
by $t$ applications of the walk $\psi(t)= A^t\psi(0)$, and by
linearity $\psi(x,y,t)$ is a linear combination of the field
$\psi(x^\prime,y^\prime,0)$ at the points lying in the past causal
cone of $(x,y,t)$.  According to Eq.~\eqref{eq:Weyl} at each time step
the field $\psi$ undergoes one of the four transitions $T_{h}$,
$h\in\{\textup{R},\textup{L},\textup{U},\textup{D}\}$, with the coin
system multiplied by the corresponding transition matrix $A_h$. A
point $(x^\prime,y^\prime,0)$ is generally connected to $(x,y,t)$ via a
number of different possible paths with the generic path conveniently
identified by a string $\sigma_t=h_t h_{t-1} \dots h_1$ of transitions
and overall transition matrix
\begin{equation}\label{eq:prod}
\mathcal{A}(\sigma_t)=A_{h_t}A_{h_{t-1}}\ldots A_{h_1}.
\end{equation}
Summing over all possible paths $\sigma_t$ and all points
$(x^\prime,y^\prime,0)$ in the past causal cone of $(x,y,t)$ one has
\begin{align}
	\label{eq:psi-paths}
	\psi(x,y,t) =\sum_{x^\prime y^\prime} \sum_{\sigma_t}
		\mathcal{A}(\sigma_t)
		  \psi(x^\prime,y^\prime,0).
\end{align}

In the following we will use the binary encoding
\begin{align}\label{eq:transition-matrices2}
A_h=\frac{1}{2} A_{ab},\quad a,b\in\{0,1\},\\
  \textup{R} \rightarrow 00, \quad \textup{L} \rightarrow 11, \quad \textup{U} \rightarrow 10, \quad
  \textup{D} \rightarrow 01,
\end{align}
with a path $\sigma_t$ uniquely identified by a $2t$-bit string
$\sigma_t=h_t h_{t-1} \dots h_1 \rightarrow s_t=a_tb_t
a_{t-1}b_{t-1}\dots a_1b_1$.

Now, in order to translate the sum over paths $\sigma_t$ in
Eq.~\eqref{eq:psi-paths} into a sum over binary strings we need a necessary and
sufficient condition that characterize all strings $s_t=a_tb_t
a_{t-1}b_{t-1}\dots a_1b_1$ that connect a pair of points
$(x^\prime,y^\prime,0)$ and $(x,y,t)$ on the causal network.  For
convenience we split the string $s_t$ in the two substrings
$\alpha_t=a_t\cdots a_1$ and $\beta_t=b_t\cdots b_1$ corresponding to
the bits in odd and even positions
\begin{align}
  s_t=(\alpha_t,\beta_t),\qquad \alpha_t,\beta_t\in\set{0,1}^{t}.
\end{align}
Upon introducing the set-bits count for the binary sub-strings
$\alpha_t$ and $\beta_t$
\begin{equation*}
    \hat{\alpha} \coloneqq \sum_{j=1}^t a_j,\quad     \hat{\beta} \coloneqq \sum_{j=1}^t b_j,
\end{equation*}
we show that given a pair of points $(x,y,t)$ and $(x',y',0)$, a
string $s_t=(\alpha_t,\beta_t)$ corresponds to a path connecting them if
and only if $t - \abs{x-x'} - \abs{y-y'}$ is even and
\begin{equation} \label{eq:cond}
\begin{split}
    \hat{\alpha} &= \frac{1}{2}(t-(x-x')+(y-y')), \\
    \hat{\beta}   &= \frac{1}{2}(t-(x-x')-(y-y')).
\end{split}
\end{equation}
These equalities are easily proved as follows. First we denote by $r,
l,u,d$ the occurrences of the
$\textup{R},\textup{L},\textup{U},\textup{D}$ transitions in the path,
with total number of steps $t = r + l + u + d$.  Now, recalling the
binary encoding $\textup{R}=00, \, \textup{L}=11, \, \textup{U}=10, \,
\textup{D}=01$, we observe that the only steps contributing to
$\hat{\alpha}$ are left and up, while the steps contributing to
$\hat{\beta}$ are left and down, namely
\begin{align*}
    \hat{\alpha} = l + u, \qquad    \hat{\beta}  = l + d.
\end{align*}
From the equations above, and noticing that $r-l = x-x'$ and $u-d =
y-y'$, one finally has
\begin{align*}
    \hat{\alpha} - \hat{\beta} &= u - d = y-y', \\
    \hat{\alpha} + \hat{\beta} &= t - (r-l) = t - (x-x'),
\end{align*}
that proves Eq.~\eqref{eq:cond}.

In simple terms a binary string $s_t=(\alpha_t,\beta_t)$ corresponds
to a path between the points $(x,y,t)$ and $(x',y',0)$ if and only if
the number of $1$-bits in the sub-strings $\alpha_t$ and $\beta_t$ are
as in \eqref{eq:cond}, and all the admissible paths are then obtained
by independent permutations of the bits in the two substrings
$\alpha_t$ and $\beta_t$. 

With the chosen binary encoding it is easy to check that the matrices
$A_{ab}$ in Eq.~\eqref{eq:transition-matrices2} generate a closed
algebra with composition rule
\begin{equation} \label{eq:2-prod}
    A_{ab} A_{cd} = (-1)^{(a \oplus c)(b \oplus d)} A_{ad},
\end{equation}
where $\oplus$ denotes the sum modulo $2$.
Accordingly, we see that the overall transition matrix \eqref{eq:prod}
associated to a path $s_t$ is given by
\begin{equation} \label{eq:gen-prod}
    \mathcal{A}(s_t) = \frac{1}{2^t}(-1)^{\phi(s_t)} A_{a_t b_1},
\end{equation}
where the phase $\phi$ can be computed by induction from its recursive
expression $\phi(s_t) = \phi(s_{t-1}) \oplus (a_{t-1} \oplus a_{t})(b_1 \oplus
b_t)$.  Starting from $\phi(s_2) = (a_1 \oplus a_2)(b_1
\oplus b_2)$ one gets the general expression
\begin{equation} \label{eq:phi-gen}
  \phi(s_t) = \bigoplus_{j \in\Integers_t} (a_{j-1} \oplus a_{j}) b_j.
\end{equation}

By Eq.~\eqref{eq:gen-prod} we see that the transition matrix of a
given path $s_t$ depends only on the first and last step, more
precisely on the first and last bits $a_t$ and $b_1$ of the string
$s_t$, and is proportional to one the four initial transition matrices
$A_{ab}$. Exploiting this feature we split the paths $s_t$ into four
equivalence classes, say $S_{ab}$ with $a,b\in\{0,1\}$, corresponding
respectively to paths having $A_{ab}$ as overall transition matrix and
the discrete path-integral of Eq.~\eqref{eq:psi-paths} is restated as
follows
\begin{equation} \label{eq:psi} 
\begin{split}
\psi(x,y,t) & = \sum_{x',y'} \sum_{s_t} \mathcal{A}(s_t)
  				\psi(x',y',0) \\
			& = \frac{1}{2^{t}}
			    \sum_{x',y'} \sum_{a,b} c_{ab} A_{ab} \psi(x',y',0), \\
  	 c_{ab} & = \! \sum_{s_t \in S_{ab}} \!\! (-1)^{\phi(s_t)},
\end{split}
\end{equation}
with the sum over all admissible paths $\sigma_t$ replaced by the sum
over all possible binary strings $s_t$. We notice that, while the
transition matrix $A_{ab}$ of a path $s_t$ depends only on the first
and last steps, the sign (plus or minus) depends in general on the
whole path according to the phase \eqref{eq:phi-gen}. As a consequence
each coefficient $c_{ab}$ is not simply the cardinality of the
equivalence class $S_{ab}$ but the sum of paths therein, each one with a sign given by its own phase.

In order to compute analytically the coefficients $c_{ab}$ of
Eq.~\eqref{eq:psi} we exploit the set-bits counts of
Eq.~\eqref{eq:cond} for paths $s_t=(\alpha_t,\beta_t)$ and the
following observation on the phase in Eq.~\eqref{eq:phi-gen}. The
phase $\phi(s_t)$ of a path $s_t=(\alpha_t,\beta_t)$ can be determined
in three steps: (i) find in the string $\alpha_t$ the number of
adjacent pairs of different bits so that $a_{j-1}\oplus a_{j}$ is not
zero, (ii) check how many of these adjacent pairs are selected by the
$1$-bits in $\beta_t$ so that $(a_{j-1}\oplus a_{j})b_j$ is not zero,
(iii) if the above number is even than the phase $\phi$ is $0$ (the
sign of the transition matrix is plus), otherwise it is $1$ (the sign
of the transition matrix is minus).

First consider the strings $\alpha_t$. For any $\alpha_t$, consecutive
equal bits can be grouped into substrings as follows
\begin{align}
\alpha_t=a_ta_{t-1}\ldots a_1=\ldots\tau^{(n)}_1\tau^{(n+1)}_0\tau^{(n+2)}_1\tau^{(n+3)}_0\ldots,
\end{align}
with $\tau^{(n)}_i=a_ia_i\dots a_i$, $i=0,1$ made of all $i$-bits. For example, consider the 7-bit string $0010111$. In this case we have $\tau^{(1)}_0=00$, $\tau^{(2)}_1=1$, $\tau^{(3)}_0=0$, $\tau^{(4)}_1=111$. Denoting by $p$ the number of $\tau_1$ slots, we see that there is a pair of different bits in correspondence to any interface---considering the string as cyclic---between slots of different type 
$\tau_1$ and $\tau_0$; consequently, any of the $p$ $\tau_1$ slots has two interfaces, 
except for the case in which $a_{t} = 1$ and $a_{1} = 1$.
Therefore, the total number of pairs of different bits is $2(p - a_t a_1)$. 
The number $u_{a a'}^p$ of strings $\alpha_t$ with 
$a_t=a,\,a_1=a'$,  with $p$ slots of type $\tau_{1}$ and $2(p - a_{t} a_{1})$ pairs of different bits is then
\begin{equation} \label{eq:n-p}
\begin{split}
	& u_{aa'}^{p} = \\
	& \begin{cases}
	  	\dbinom{\hat\alpha-1}{p-1}
		\dbinom{t-\hat\alpha-1}{p-a-a'}, & \text{if $0 < \hat\alpha < t$,} \\ \\
		1,			  				     & \text{if $\hat\alpha = t \land a a' = 1$ or} \\
									     & \hphantom{if } \hat\alpha = 0 \land (1\oplus a)(1\oplus  a') = 1, \\
		0,			  					 & \text{if $\hat\alpha = t \land a a' =0 $ or} \\
										 & \hphantom{if } \hat\alpha = 0 \land (1\oplus a)(1\oplus  a')=  0.
	  \end{cases}
\end{split}
\end{equation}
%
\iffalse
\begin{align} \label{eq:n-p} 
&u_{a a'}^p=\\\nonumber
&\begin{cases}
\PMbinom{\hat\alpha}{p-1}
\PMbinom{t-\hat\alpha-1}{p-a-a'},\qquad \qquad\:\:\:\: \text{if}\:\:\: 0<\hat\alpha <t,\\
0,\qquad \text{if}\:\:\: \hat{\alpha}=t, a+ a'< 2\:\:\:\text{or}\:\:\: \hat\alpha=0, a+ a'> 0,\\
1, \qquad\text{if}\:\:\: \hat{\alpha}=t, a+ a'= 2\:\:\:\text{or}\:\:\: \hat\alpha=0, a+ a'=0,
\end{cases}
\end{align}
\fi
%
To prove Eq.~\eqref{eq:n-p} we remind that in $\alpha_t$ there are
$\hat\alpha$ 1-bits and $t-\hat\alpha$ 0-bits that must be arranged
independently in $p$ and $p-a-a'+1$ slots respectively. The number of
these arrangements is given by the product of the $p$-compositions of
$\hat\alpha$ with the $(p-a-a'+1)$-compositions of $t-\hat\alpha$ (the
number of $p$-compositions of an integer $n$ is
$\SMbinom{n-1}{p-1}$), which gives Eq.~\eqref{eq:n-p}.

Now we consider the strings $\beta_t$. Given the strings $\alpha_t$ with
$a_t=a,\,a_1=a'$ and $\mu=2p-a-a'$ free pairs of different bits
\footnote{ Since the total number of pairs of different bits is $2(p -
  aa')$, and the last and first bits $a_t=a,\,a_1=a'$, $p\,
  \tau_{1}$ in $\alpha_t$ are fixed, the number of free pairs is $\mu
  \coloneqq 2(p - aa') - a \oplus a' = 2p - a - a'$.}, the number of strings
$\beta_t$ with $b_1=b$ that select $k+b(a\oplus a')$ pairs in $\alpha_t$ is denoted
$w_{aa'b}^{p,k}$ and is given by
\begin{equation} \label{eq:c-p-k}
\begin{split}
	& w_{aa'b}^{p,k} = \\
	& \begin{cases}
		\dbinom{\mu}{k} \dbinom{t-\mu-1}{\hat{\beta}-k-b}, &
                \text{if $0 < \hat\beta < t$,} \\
		1,												   & \text{if $\hat\beta = 0 \land b = 0$ or} \\
														   & \hphantom{if } \hat\beta = t \land b = 1, \\
   	 	0,												   & \text{if $\hat\beta = 0 \land b = 1$ or} \\
														   & \hphantom{if } \text{$\hat\beta = t \land b = 0$}. \\
	  \end{cases}
\end{split}
\end{equation}
Indeed among the $\hat\beta - b$ free $1$-bits of $\beta_{t}$, one
uses $k$ of them to select $k$ free pairs, which can be done in
$\SMbinom{\mu}{k}$ ways.  The remaining $\hat\beta-k-b$
%(the first bit $b_1=b$ is already fixed) 
1-bits must be arranged in $t-\mu-1$ places, which is done in
$\SMbinom{t-\mu-1}{\hat\beta-k-b}$ ways.  This proves
Eq.~\eqref{eq:c-p-k}. 

Finally we can calculate the coefficients $c_{ab}$ as
\begin{equation} \label{eq:coefficients1}
	c_{ab} = \sum_{p=p\Tped{min}}^{p\Tped{max}} \sum_{a'=0,1} \sum_{k=k\Tped{min}}^{k\Tped{max}} 
				(-1)^{k+b(a \oplus a')} u_{a a'}^{p} w_{aa'b}^{p,k},
\end{equation}
with
\begin{align*}
	&p\Tped{min} = \min(1, \hat\alpha),\quad p\Tped{max}  = \min(\hat\alpha, t - \hat\alpha), \\
	&k\Tped{min} = \max(0, \hat\beta + \mu - t - b), \quad k\Tped{max} = \min\paren*{\abs{\mu}, |\hat\beta - b|}.
\end{align*}
where the product $\sum_a'u_{aa}^{p} w_{aa'b}^{p,k}$ is the total
number of paths with: last bit $a_t=a$ and first bit $b_1=b$,
$2(p-aa')$ pairs of different bits in $\alpha_t$ among which
$k+b(a\oplus a')$ selected by the 1-bits in $\beta_t$.

It is worth noticing that the sign, plus or minus, is decided by the total number of pairs
$k + b(a \oplus a')$ selected with the strings $\beta_t$. 
%Finally, in order to consider all the possible paths, one has to sum over all the possible values of $k$ and $p$.
For $p>1$ and $0 < \hat\beta < t$, the sum over $k$ in Eq.~\eqref{eq:coefficients1} can be evaluated in
terms of the hypergeometric function ${}_2F_1(a,b,c,z)$,
considering that one can extend the sum over all $\Naturals$ 
since the binomials $\binom{n}{k}$, with $n \in \Naturals$, are null 
%if $k \in \set{z \in \Integers | z < 0 \lor z > n}$:
for all $k \in \Integers, \, k < 0 \text{ or } k > n$:
\begin{align} \label{eq:c-p}
	\tilde{w}^{p}_{aa'b} & \coloneqq \sum_{k=0}^{+\infty} (-1)^{k + b(a\oplus a')} 
								\binom{\mu}{k} \binom{t - \mu - 1}{\hat\beta - k - b} \nonumber \\
        				 & \,= (-1)^{b(a \oplus a')}
				 				\binom{t - \mu - 1}{\hat\beta - b} F_{aa'b}^p, \\
    F_{aa'b}^p			 & \coloneqq {}_2 F_1(b - \hat\beta, -\mu, t - \hat\beta - \mu + b, -1). \nonumber
\end{align}
In this way we have a simplified expression for the coefficients $c_{ab}$:
\begin{equation} \label{eq:cof}
    c_{ab} = \sum_{p=p\Tped{min}}^{p\Tped{max}} \sum_{a'=0,1}
                    u_{aa'}^p \tilde{w}_{aa'b}^p.
\end{equation}

In this Letter we presented a method for the expression of a QW via a path-sum. The procedure is based on the binary encoding of the walk paths and on the closure of the transition matrices algebra. Using this approach we provided an analytical solution for the Weyl QW in two space dimensions, providing the first example of discrete path-integral solution for a QW in dimension higher than one.

The technique used in this paper could in principle be generalized to any QW whose transition matrices generate a closed algebra under multiplication. It is then interesting to investigate which are the hypotheses on the dynamics of the walk that imply this simple algebraic feature. One can conjecture that the closure of the transition matrices algebra is a consequence of the QW's symmetries. Indeed both the Weyl and the Dirac QWs of Ref.~\cite{d2013derivation}, which exhibit this remarkable property, are covariant with respect to the discrete symmetries of the underlying lattice, while removing the covariance requirement introduces QWs that do not enjoy the closure property.

On the other hand, it is not clear to what extent one can generalize the other two conditions that allow for the analytic computation, namely the i) possibility of classifying paths connecting fixed vertices with the same transition matrices in terms of a suitable encoding of elementary transitions, and ii) the possibility of calculating the complex amplitude of paths as a simple function of the encoding.

\section{Acknowledgments}

This work has been supported in part by the Templeton Foundation
under the project ID\# 43796 A Quantum-Digital Universe.

%%%%%%%%%%%%%%%%%%%%%%%%%%%%%%%%%%%%%%%%%%%%%%%%%%%%%%%%%%%%%%%%%%%%%%%%%%%%%%%%%%%%%%%%%%%%%
\bibliographystyle{eplbib}
\bibliography{bibliography}
%%%%%%%%%%%%%%%%%%%%%%%%%%%%%%%%%%%%%%%%%%%%%%%%%%%%%%%%%%%%%%%%%%%%%%%%%%%%%%%%%%%%%%%%%%%%%

\end{document}